\documentstyle[prl,aps,epsf]{revtex}
\draft
\begin{document}
\twocolumn[\hsize\textwidth\columnwidth\hsize\csname @twocolumnfalse\endcsname

\title{Hysteretic Depinning and Dynamical Melting for Magnetically Interacting 
Vortices in Disordered Layered Superconductors}
\author{C. J. Olson and C. Reichhardt}
\address{Department of Physics, University of California, Davis, California
95616}

\author{V. M. Vinokur}
\address{Materials Science Division, Argonne National Laboratory, 
Argonne, Illinois 60439}

\date{\today}
\maketitle
\begin{abstract} 
We examine the depinning transitions and the temperature versus
driving force phase diagram for magnetically interacting pancake vortices
in layered superconductors.  
For strong disorder the initial depinning is plastic
followed by a sharp hysteretic transition to a 3D ordered state for increasing 
driving force. 
Our results are in good agreement with theoretical predictions 
for driven anisotropic charge density wave systems. 
We also show that a temperature induced peak
effect in the critical current
occurs due to the onset of plasticity between the layers.  
\end{abstract} 

\vskip2pc]
\narrowtext
A vortex lattice subject to quenched disorder offers an ideal system 
for studying glassy dynamics governed by the competition between the 
ordering forces (the vortex-vortex repulsion) and disorder \cite{Blatter1}.
One of the key issues of glassy dynamics is the nature of the {\it 
depinning transition} separating a pinned state of the lattice at small 
external drives from the sliding regime that occurs when
the external force exceeds a threshold depinning value.  The early numerical 
simulations revealed that in the system with strong 
disorder plastic depinning occurs
with the vortices tearing past one 
another and flowing in complex patterns \cite{Jensen2}.
The onset of plasticity at depinning was later confirmed in 
experiment \cite{Higgins3} as well as subsequent numerical studies 
\cite{Vinokur4,Num5} and was demonstrated to
play a crucial role in 
the pre-melting peak effect where the
critical current exhibits a striking increase 
as a function of temperature or field \cite{Blatter1,Higgins3} prior to 
vortex lattice melting and 
related phenomena. 
As the applied driving force is further increased,
the effect of the quenched disorder is reduced and the vortex 
lattice can show a remarkable reordering 
transition at a driving force $F_{r}$ \cite{Vinokur4}. 
At this transition a considerable portion of the vortex lattice
order is restored and the motion is highly coherent.
The reordering force $F_{r}$ is expected to
diverge as the temperature $T$ approaches 
the vortex lattice melting temperature $T_{m}$ \cite{Vinokur4}. 
Transport measurements provide
strong evidence for this scenario \cite{Higgins3}.
Additional studies on the 
depinning and reordering transition and the nature of the
coherently moving phases have been 
conducted theoretically \cite{Reorder6}, 
numerically \cite{Moon7}, and experimentally \cite{Kes8,Andrei9,Fendrich10}, 
showing both  
non-hysteretic continuous depinning transitions  
and abrupt hysteretic V(I) curves.
The appearance of the hysteretic 
portion of the V(I) curves depends on how the vortex
lattice was prepared: the initial current ramp shows a large 
hysteretic critical current but subsequent ramps produce
only continuous non-hysteretic V(I) characteristics \cite{Andrei9}.  

The underlying mechanisms of depinning and its interplay with the 
reordering transition remain an open question. The issues to be 
clarified include 
a determination of the conditions under which
the depinning transition shows hysteretic or 
switching responses, as well as an understanding
of the nature of these phenomena.  
The mechanism by which a three-dimensional 
lattice depins or reorders is also an open question,
as well as the issue of whether 
the onset of plasticity coincides with a sudden jump in the
critical current (the peak effect).

Some of these issues were addressed in \cite{Nattermann11} in a context 
of  a 
driven three-dimensional anisotropic charge density 
wave system (CDW) with disorder.  It was shown that if disorder is 
sufficiently strong\cite{note12}, the dependence of the lattice 
velocity on the applied drive becomes {\it two}-valued or bistable: 
at the same 
applied force (close to the depinning force of the topologically 
disordered phase) the lattice can either remain at rest or slide 
depending on the history.  This means that the depinning transition 
shows switching behavior and hysteresis.  The details of depinning are 
determined by particulars of the system involved.  For a layered
anisotropic CDW considered in  \cite{Nattermann11} the depinning occurs 
in two stages.  The initial plastic depinning, where the decoupled 
2D CDWs slide independently in each layer, is followed 
by
a sharp transition or crossover at higher drives to a 
faster coherently moving 
3D solid CDW phase.  Reversing the process reveals hysteresis:
pinning (immobilizing) of a coherent 3D CDW occurs at {\it smaller} 
drives. 
It was also shown that for  
weak disorder there is only a
single non-hysteretic depinning 
transition directly to the 3D coherently moving state.  

The results of 
\cite{Nattermann11} are very general, and the particular model used there 
can be straightforwardly generalized
to other periodic media in quenched disorder.
A detailed study of the character of the many-valued $v-F$ dependence
near depinning was recently
carried out in \cite{Marchetti12a} within the framework of a mean-field model
for a visco-elastically driven vortex configuration, where a dynamic
bistability and, accordingly, a hysteretic plastic depinning was
found to occur for sufficiently strong disorder.
Note that the layered CDW model
of \cite{Nattermann11} can be conveniently
applied directly to
highly anisotropic superconductors where  
the magnetic interactions between pancake vortices
dominate \cite{clem13}.  
In this system both a 2D regime, in which vortices in each layer move
independently from the other layers, and a 3D regime, where vortices in
different layers form coherently moving lines, should exist
\cite{3D14,kolton14a}. 

In this work we investigate the dynamics of magnetically interacting vortices
in a three-dimensional layered superconductor to numerically test the 
predictions of 
Ref. \cite{Nattermann11}.  
Our model is relevant to anisotropic layered superconductors such 
as BSCCO as well as artificially layered superconductors in which
the magnetic interactions dominate. Recent numerical simulations 
with this model have found that for increasing applied driving force,  
a transition from a decoupled plastically flowing state
to a coupled 3D coherently moving lattice occurs \cite{3D14,kolton14a}.
A sharp static 2D to 3D disorder induced transition also appears as a  
function of field and disorder \cite{3D14,OurPeak14b}. 
In this model,
the dynamic phase diagram as a function of driving
force and temperature, as well as hysteretic and 
switching behavior in the current-voltage characteristics, have 
not been studied until now. 
We find that for strong disorder, the initial
depinning is 2D with the vortices flowing plastically and independently 
in any one layer.
For increased drives we see a sharp transition to a coherently moving 
3D vortex lattice 
which manifests itself in an abrupt increase 
in the vortex velocities. 
We observe strong hysteresis when the driving current is cycled.
We have also investigated the driving versus temperature phase diagram.
 
We consider a three-dimensional layered superconductor in which
the vortices are modeled as repulsive 
point particles confined to a layer with
an equal number of vortices per layer.
The overdamped equation of motion for vortex $i$ is given by
$$ {\bf f}_{i} = -\sum_{j=1}^{N_{v}}\nabla_{i} {\bf U}(\rho_{i,j},z_{i,j})
+ {\bf f}_{i}^{vp} + {\bf f}_{d} + {\bf f}^{T}= \eta{\bf v}_{i} \ ,$$
where $N_v$ is the number of vortices, $\rho_{i,j}$ and 
$z_{i,j}$ are the distance
between pancakes $i$ and $j$ in cylindrical coordinates, and $\eta=1$.
The system has periodic boundaries in-plane and open boundaries
in the $z$ direction \cite{ngj15}.
The magnetic energy between pancakes is \cite{clem13,brandt16}
\begin{eqnarray}
{\bf U}(\rho_{i,j},0)=2d\epsilon_{0} 
\left((1-\frac{d}{2\lambda})\ln{\frac{R}{\rho}}
+\frac{d}{2\lambda} 
E_{1}\right) 
\nonumber
\end{eqnarray}
\begin{eqnarray}
{\bf U}(\rho_{i,j},z)=-s_{m}\frac{d^{2}\epsilon_{0}}{\lambda}
\left(\exp(-z/\lambda)\ln\frac{R}{\rho}- 
E_{2}\right) \nonumber
\end{eqnarray}
where
$\epsilon_{0} = \Phi_{0}^{2}/(4\pi\lambda)^{2}$, 
$\lambda$ is the London penetration depth,
$R = 22.6 \lambda$, the maximum radial distance,
$E_{1} = 
\int^{\infty}_{\rho} d\rho^{\prime} \exp(-\rho^{\prime}/\lambda)/\rho^{\prime}$,
$E_{2} = 
\int^{\infty}_{\rho} d\rho^{\prime} \exp(-\sqrt{z^2 + \rho^{\prime 2}}/\lambda)/\rho^{\prime}$,
$d=0.005\lambda$ is the interlayer spacing,
and we set $s_m=2.0$.
The uncorrelated pins are modeled by parabolic traps that are
randomly distributed in each layer.  The vortex-pin interaction is given by
${\bf f}_{i}^{vp} = -\sum_{k=1}^{N_{p,L}} (f_{p}/\xi_{p})
({\bf r}_{i} - {\bf r}_{k}^{(p)}) \Theta (
(\xi_{p} - |{\bf r}_{i} - {\bf r}_{k}^{(p)} |)/\lambda)$,
where the pin radius $\xi_{p}=0.2\lambda$, the
pinning force $f_{p}=0.02f_{0}^{*}$, and $f_{0}^{*}=\epsilon_{0}/\lambda$.
We have simulated a $16\lambda \times 16\lambda$ system 
with a vortex density of $n_v = 0.35/\lambda^2$ and a pin
density of $n_p = 1.0/\lambda^2$ in each of $L=8$ layers.  This
corresponds to $N_{v}=80$ vortices and $N_{p}=256$ pins per layer, with
a total of 640 pancake vortices. 
For our finite temperature simulations the temperature is 
modeled as a Langevan kick ${\bf f}^{T}$ where 
$<{\bf f}(t)^{T}> = 0$ and $<{\bf f}(t)^{T}_{i}{\bf f}(t)^{T}> 
= 2\eta k_{B}T\delta_{ij}$. 

In Fig.~1(a) we show the average vortex velocities in the
direction of drive, $V_{x}=<v_{x}>$, as the applied driving force
$f_{d}$ is increased and then decreased, 
and in Fig.~1(b) we show the corresponding c-axis correlation function, 
$C_z = 1 - \langle(|({\bf r}_{i,L} - {\bf r}_{i,L+1})|/(a_0/2))
\Theta( a_0/2 - |({\bf r}_{i,L} - {\bf r}_{i,L+1})|)\rangle$, where
$a_0$ is the vortex lattice constant. 
At $f_d=0$ the vortices
are pinned and uncorrelated in the $z$-direction as is indicated by
the fact that $C_z$ has a low value.
The initial depinning, at $f_{dp} = 0.01f_{0}^{*}$, 
is 2D with the vortices remaining 
uncorrelated in the $z$-direction.  Near $f_{d} = 0.013$
the vortex velocities sharply increase, coinciding with a sharp 
transition to a coupled moving 3D state as seen in the jump up in 
$C_z$ \cite{3D14}.  At higher drives
the vortex velocity linearly increases with increasing 
$f_{d}$. Upon reducing $f_{d}$ from this linear regime the system remains 
in the 3D state for driving forces well below $f_d=0.013$.  At
$f_d=0.007$
there is a sudden transition to a 2D pinned state as 
seen in the drop in $V_x$ and $C_z$. If the applied driving force is again
increased 
the velocities and $C_z$ will follow the same hysteresis loop. 
In Fig.~1(c,d) we show $V_x$ and $C_z$ for the same system as in 
Fig.~1(a,b) but with a lower disorder strength of $f_{p} = 0.01f_{0}^{*}$. 
Here the depinning transition is
non-hysteretic with a single 
depinning threshold to the 3D coherently moving state as seen by the
fact that $C_z = 1.0$ at all times. 
For a given coupling strength value $s_m$, 
there is a critical pinning force above which there is
a static 3D-2D transition \cite{3D14}.  Hysteretic $V(I)$ curves can only be 
observed above, but close to, this transition. 
The width 
of the hysteresis in the $V(I)$ curves is largest for 
disorder strengths just above the static 2D-3D transition and gradually 
narrows for increasing disorder strength.   

The behavior we observe 
is in excellent agreement with the theoretical predictions
of \cite{Nattermann11}.
We note that unlike the
anisotropic charge-density wave system, where only inter-plane
plasticity or slipping can occur,
in the vortex system in-plane plasticity is possible within each
layer. 
The 2D-3D transition in the case of the vortices occurs by 
simultaneous recoupling of the vortices between layers
and the reordering of the vortices in plane. 
To illustrate this, in Fig.~2(a) we show a top
view of the vortices in 

\begin{figure}
\centerline{
\epsfxsize=3.5in
\epsfbox{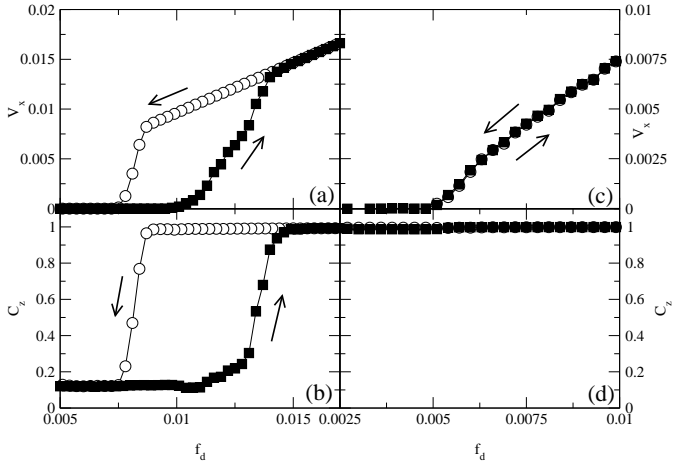}}
\caption{(a) Vortex velocities $V_{x}$ versus $f_{d}$ for a system 
with $f_{p}=0.02f_{0}^{*}$ at $T=0$ 
as the driving force is first increased from 
$f_{d} = 0.0$ to $f_{d} = 0.02$ and then decreased back to $f_{d} = 0$. 
(b) $C_z$ versus $f_{d}$ for the system in (a) showing the sharp,
hysteretic recoupling and decoupling transitions.
(c) $V_{x}$ versus $f_{d}$ for a system with $f_{p} = 0.01f_{0}$. 
(d) $C_z$  versus $f_{d}$ for the system in (c) showing
that there is a non-hysteretic 3D pinned to 3D moving transition. 
}
\label{fig1}
\end{figure} 

\hspace{-13pt}
the moving 2D regime, 
indicating that the vortices are
uncorrelated from layer to layer and are disordered in plane. 
In contrast, Fig.~2(b) shows that in the 
moving 3D regime, the vortices are aligned 
between layers and possess a triangular ordering.  

In Fig.~3 we show the driving force versus temperature phase diagram for 
a system with $f_{p}=0.01f_{0}^{*}$.
The temperature is plotted in units of $T_{m}=0.00045$, the temperature
at which the clean system undergoes melting in the form of
a single sharp transition from a 3D lattice to a 2D pancake gas.
As shown in Fig. 3, at $T/T_{m}<0.005$ for low drives there is a
3D pinned phase where the vortices remain aligned.  At these temperatures,
for $f_{d}/f_{0}=0.001$ there is a non-hysteretic depinning transition 
directly to the coherently moving 3D state. 

We find a static transition to the 2D state 
at $T/T_{m} = 0.005$, which we label $T_{dc}$. 
For $T>T_{dc}$ the depinning transition is 2D, and as the driving
force is further increased a dynamically induced reordering of
the vortices occurs.
As $T$ is further increased above $T_{dc}$  
the size of the 2D pinned phase is reduced, while the drive at
which the
2D plastic flow to 3D coherent flow transition occurs
diverges as $T$ approaches $T_{m}$, in agreement with
the dynamical freezing model of Koshelev and Vinokur \cite{Vinokur4}. 

The thermally induced decoupling seen in Fig. 3
also coincides with a sharp increase in the critical current 
or a ``peak effect.''
There has been considerable work on dimensional transitions in the
context of the peak effect in layered
superconductors.  The static 3D-2D transition 
and critical current increase in layered
superconductors have been previously studied as a function of interlayer 
coupling and magnetic field \cite{Peak17,Peak218}. 
Koshelev and Kes 

\begin{figure}
\centerline{
\epsfxsize=3.5in
\epsfbox{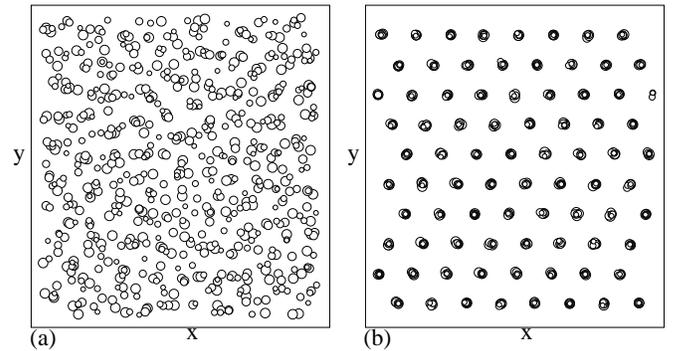}}
\caption{Pancake vortex positions seen from the top of the sample,
for the system with $f_{p}=0.02f_{0}$ shown in Fig. 1(a,b).
Pancakes on a given layer are represented by circles with a fixed
radius; the radii increase from the top layer to the bottom.  Vortices
that have aligned into vortex lines appear as thick circles.
(a) At $f_{d}=0.011f_{0}^{*}$, in the 2D plastic flow regime.
(b) At $f_{d} = 0.015f_{0}^{*}$ in the coherently moving 3D phase. 
}
\label{fig2}
\end{figure}  

\hspace{-13pt}
have proposed
theoretically that magnetically coupled vortices can show a
first order transition from a 3D to a 2D system 
under certain conditions \cite{KesPeak19}. 
The sharpness in the 3D-2D transition we observe in our simulation
suggests that it is first order in nature \cite{3D14,OurPeak14b}.

The 3D-2D transition can be seen as a temperature induced peak 
effect in which the combination of temperature and
quenched disorder bring about a decoupling transition where the 
individual pancakes can adjust to the optimal pinning configurations. 
The phase diagram in Fig.~3 is very similar to the one proposed
by Bhattacharya and Higgins \cite{Higgins3} 
in which there is an elastic depinning
regime for a certain temperature range, above which the peak effect and 
the onset of plasticity simultaneously appear. 
In our model the peak effect is caused not by the onset of plasticity
in the $a - b$ plane but by the
onset of  plasticity in the $c$-axis \cite{OurPeak14b} with the plasticity only
occurring for $T > T_{dc}$. 
We find that the temperature $T_{dc}$ moves further below $T_{m}$ 
as the disorder in the sample is increased. 
Thus for large enough disorder the 2D phase extends all the way
to $T = 0$, and the dynamic phase diagram becomes the same as the
one proposed in in Ref. \cite{Vinokur4}.

In Fig.~4 we show the 
$V_x$ versus $f_{d}$ curves, which correspond to experimental V(I) curves, 
for increasing $T$. For 
$T< T_{dc}$  there is a sharp elastic depinning transition
at which all the vortices start moving at once in the 
3D state. 
For $T > T_{dc}$ the depinning response is more continuous since
the 2D depinning occurs inhomogeneously, with certain regions moving while
other regions remain pinned. As $T$ is further increased in the 2D regime
the depinning transition falls at driving forces lower than that 
at which the 3D depinning transition occurred.
Although for these higher temperatures the 2D pancakes depin 
at a lower applied drive than the 3D state, 
the overall velocity above depinning is still less than
that of a coherently moving 3D system at the same drive so that 
a crossing of the IV curves occurs. 

To summarize, we have numerically 
investigated the dynamics of magnetically coupled
vortices in layered superconductors interacting with quenched 
disorder. 
For sufficiently strong disorder the depinning is 2D.  Here the vortices 
flow plastically in each layer and are uncorrelated from layer
to layer. For
higher drives there is a sharp hysteretic transition to a 
coherently moving ordered
3D vortex lattice. For weak disorder there is a non-hysteretic 
depinning transition from a pinned 3D to a moving 3D state. The two stage
2D-3D hysteretic depinning transition is in good agreement with the theoretical
predictions of Ref. \cite{Nattermann11}.  
We have also mapped out the dynamic phase diagram 
as a function of driving force and temperature. At low temperatures we
observe non-hysteretic 3D pinned to 3D moving transitions. For
increased temperature there is a static 3D-2D transition which coincides 
with a peak in the critical current as well as the onset of plasticity in 
the $c$-axis. The drive at which a dynamically induced 
2D-3D dynamic transition diverges as $T_{m}$ is approached. 
Our dynamic phase 
diagram is very similar to that proposed in Ref\cite{Higgins3}.  

CJO and CR thank B. Janko for his kind hospitality at Argonne National 
Laboratory.  This work was supported
by CLC and CULAR (LANL/UC), by NSF-DMR-9985978, and 
by Argonne National Laboratory through
U.S.\ Department of Energy, Office of Science under contract
No.~W-31-109-ENG-38.

\begin{figure}
\centerline{
\epsfxsize=3.5in
\epsfbox{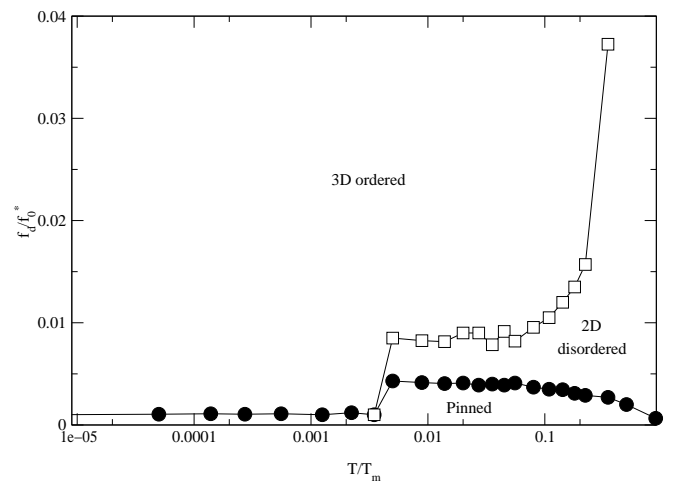}}
\caption{The $f_{d}$ versus $T/T_{m}$ dynamic phase diagram for a system
with $f_{p}=0.01f_{0}^{*}$. 
Filled circles: Depinning line.  Open squares: Dynamic reordering line.
}  
\label{fig3}
\end{figure}

\begin{figure}
\centerline{
\epsfxsize=3.5in
\epsfbox{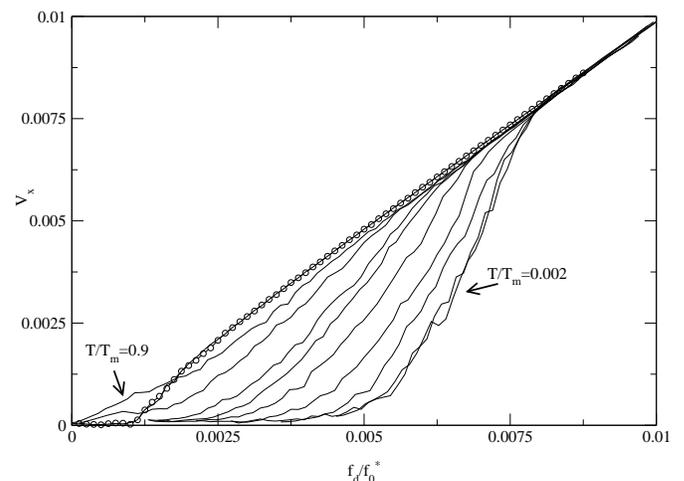}}
\caption{
$V_{x}$ versus $f_{d}$ responses for varied $T$.  Circles: temperatures below
the decoupling temperature $T_{dc}$: $T/T_{m}=0.00027$ and $0.0013$.
Lines, right to left: $T/T_{m}=$ 0.005, 0.014, 0.05, 0.11, 0.22,
0.5, 0.9.
}
\label{fig4}
\end{figure} 

\end{document}